\documentclass[12pt]{article}

\usepackage[latin1]{inputenc}
\usepackage[spanish,english]{babel}
\usepackage{amsfonts}
\usepackage{amssymb}
\usepackage{graphicx}
\newcommand{\be}{\begin{equation}}
\newcommand{\ee}{\end{equation}}
\newcommand{\bea}{\begin{eqnarray}}
\newcommand{\eea}{\end{eqnarray}}

\begin{document}
\begin{titlepage}

\begin{flushright}
{\tt
  hep-th/0701162}
 \end{flushright}

\bigskip

\begin{center}
{\Large \bf Short distances, black holes, and TeV
gravity\footnote{To appear in the proceedings of the Eleventh
Marcel Grossmann Meeting (Berlin, July 2006)}}
\\\vspace{1cm}

I. Agullo$^{a}$\footnote{ivan.agullo@uv.es},
 J. Navarro-Salas$^a$\footnote{jnavarro@ific.uv.es} and  Gonzalo J.
 Olmo$^{b}$\footnote{olmoalba@uwm.edu}
\end{center}
\bigskip

\footnotesize \noindent {\it a) Departamento de F\'{\i}sica
Te\'orica and
    IFIC, Centro Mixto Universidad de Valencia-CSIC.
    Facultad de F\'{\i}sica, Universidad de Valencia,
        Burjassot-46100, Valencia, Spain.}\\
{\it c) Physics Department, University of Wisconsin-Milwaukee,
P.O.Box 413, Milwaukee, WI 53201 USA USA}

\bigskip

\bigskip

\begin{center}
{\bf Abstract}
\end{center}

The Hawking effect  can be rederived in terms of two-point functions
and in such a way that it makes it possible to estimate, within the
conventional semiclassical theory, the contribution of ultrashort
distances at $I^+$ to the Planckian spectrum. Thermality is
preserved for black holes with $\kappa l_P \ll1$. However,
deviations from the Planckian spectrum can be found for mini black
holes in TeV gravity scenarios, even before reaching the Planck
phase.

\bigskip

\end{titlepage}
In 1974 Hawking predicted the thermal emission of quanta by black
holes \cite{hawk1} using semiclassical gravity . The deep connection
of this result with thermodynamics and, in particular, with the
generalized second law, strongly support its robustness and its
interpretation as a low-energy effect, not affected by the
particular underlying theory of quantum gravity \cite{books} .
However\cite{jacobson9193} , ultrahigh frequencies (or ultrashort
distances) seem to play a crucial role in the derivation of the
Hawking effect. Any emitted quanta, even those with very low
frequency at future infinity, will suffer a divergent blueshift when
propagated backwards in time and measured by a freely falling
observer. The exponential redshift of the event horizon provides, to
the external observer, a glimpse of the world at very short-distance
scales, where semiclassical tools are not well justified. All
derivations of Hawking radiation seem to invoke Planck-scale physics
in a fundamental way, which makes it unclear the way to parameterize
the contribution of transplanckian physics in black hole radiation.

 We propose an alternative to the standard
approach in terms of Bogolubov coefficients to derive the Hawking
effect. In our approach, the correlation functions of the matter
fields are used to compute the spectrum of the emitted particles.
 This provides an explicit way to evaluate the contribution of ultrashort  distances (Planck-scale) to the
 spectrum of Hawking quanta within the semiclassical approach.

Let us assume, for the sake of simplicity, that $\phi$ is a massles,
neutral, and minimally coupled scalar field. One can easily verify
that the number operator can be obtained from the following
projection
\begin{equation}\label{eq:ai+aj}
{a^{out}}^\dagger_i a^{out}_j= \int_\Sigma d\Sigma_1 ^\mu d\Sigma_2
^\nu [u^{out}_i(x_1){\buildrel\leftrightarrow\over{\partial}}_\mu
][u^{out*}_j(x_2){\buildrel\leftrightarrow\over{\partial}}_\nu ]
 \left\{\phi (x_1)\phi (x_2)- \langle out|\phi (x_1)\phi
(x_2)|out\rangle\right\}
 \ , \ \ \ \end{equation}
where $u^{out}_i(x)$ is a normalized positive frequency mode with
respect to the inertial time at future infinity, and $\Sigma $
represents a suitable initial value hypersurface. Therefore, the
number of particles in the $i^{th}$ mode measured by the ``out''
observer in the ``in'' vacuum is given by $\langle
in|N_i^{out}|in\rangle \equiv \langle in|N_{ii}^{out}|in\rangle$,
where $N_{ij}^{out}\equiv\hbar^{-1}{a^{out}}^\dagger_i a^{out}_j$
can be easily worked out using the above expression. Let us now
apply (\ref{eq:ai+aj}) to the formation process of a Schwarzschild
black hole and restrict the ``out'' region to future null infinity
($I^+$). The ``in'' region is, as usual, defined by past null
infinity ($I^-$). At $I^+$ we can consider the normalized radial
plane-wave modes $ u^{out}_{wlm}(t,r, \theta, \phi) =
u_{w}(u)r^{-1}Y_{lm}(\theta, \phi)$, where $u_{w}(u)=
\frac{e^{-iwu}}{\sqrt{4\pi w}}$ and $u$ is the null retarded time.
Using these modes in  (\ref{eq:ai+aj}) one
finds\cite{agullo-navarro-salas-olmo06,agullo-navarro-salas-olmo-parker}
\ (for simplicity we omit the factor $\delta_{l_1 l_2}\delta_{m_1
m_2}$)

\begin{equation}\label{eq:divergences}
\langle in|N_{i_1 i_2}^{out}|in\rangle = -\frac {|t_{lm}(w)|^2\delta
(w_1-w_2)}{2\pi\sqrt{w_1w_2}} \int_{-\infty}^{+\infty} dz
e^{-i\frac{(w_1+w_2)}{2}z} \left[ \frac{\kappa^2 e^{-\kappa
z}}{(e^{-\kappa z}-1)^2} -\frac{1}{z^2} \right] \nonumber \
\end{equation}
where $z=u_1-u_2$ represents the ``distance'' between the points
$u_1$ and $u_2$ and $t_{lm}(w)$ are the transmission coefficients.
To get the Planckian spectrum, there remains to perform the
integration in $z$
\begin{equation}\langle in|N_{w}^{out}|in\rangle=\frac{-|t_{lm}(w)|^2}{2\pi w}
\int_{-\infty}^{+\infty} dz e^{-iwz} \left[ \frac{\kappa^2
e^{-\kappa z}}{(e^{-\kappa z}-1)^2} -\frac{1}{z^2}\right] =
\frac{|t_{lm}(w)|^2}{e^{2\pi w \kappa^{-1}}-1}\nonumber \ .
\end{equation} The interesting aspect of the above expression is
that it allows us to explicitly evaluate the contribution of
distances to the thermal spectrum. To be more explicit, the integral
\begin{equation}\label{franja}
I(w,\kappa,\epsilon)=- \frac{1}{2\pi w} \int_{-\epsilon}^{+\epsilon}
dz e^{-iwz} \left[ \frac{\kappa^2 e^{-\kappa z}}{(e^{-\kappa
z}-1)^2} -\frac{1}{z^2}\right] \nonumber
\end{equation}
can be regarded as the contribution coming from distances $z \in
[-\epsilon, \epsilon]$ to the full spectrum. This integral can be
solved analytically. For details and the case of a massless spin
$s=1/2$ field see
\cite{agullo-navarro-salas-olmo06,agullo-navarro-salas-olmo-parker}
.
Obviously, in the limit $\epsilon\to \infty$, we recover the
Planckian result $I(w,\kappa,\infty)=(e^{2\pi
w\kappa^{-1}}-1)^{-1}$. For a rotating black hole the result is
similar with the usual replacement of $w$ by $\tilde{w}\equiv
w-m\Omega_H$ ($m$ is the axial angular momentum quantum number of
the emitted particle and $\Omega_H$ the angular velocity of the
horizon).

On the other hand, if we take $\epsilon$ of order of the Planck
length $l_P=1.6\times10^{-33}cm$, we obtain that the contribution to
the thermal spectrum at the typical emission frequency,
$w_{typical}\sim\kappa/2\pi\equiv T_H$, due to transplanckian scales
is of order $\kappa l_P$. This contribution is negligible for
macroscopic black holes with typical size much bigger than the
microscopic Planck length. In fact, for three solar masses black
holes the contribution to the total spectrum, $(e^{2\pi
w\kappa^{-1}}-1)^{-1}$, at $w_{typical}$ is of order $10^{-38}\%$.
We need to look at high frequencies, $w/w_{typical}\approx 96$, to
get contributions of the same order as the total spectrum itself.
This is why Hawking thermal radiation is very robust, as it has been
confirmed in analysis based
on acoustic black holes\cite{unruh95}\ .\\
Our results, in addition, indicate that when the product $\kappa
l_P$ is of order unity, the contribution of short distances to the
Planckian spectrum is not negligible. The integral $I(w,\kappa,
\epsilon)$ gives values similar to $(e^{2\pi w \kappa^{-1}}-1)^{-1}$
when $w/w_{typical}$ is not very high. This happens in the case of
black holes predicted by TeV gravity scenarios
\cite{extradimensions,giddings-thomas02} . For detailed and recent
results see \cite{casals} . Assuming a drastic change of the
strength of gravity at short distances due to $n$ extra dimensions
(a Planck mass $M_{TeV}$ of 1 TeV) and for a $(4+n)$-dimensional
Kerr black hole with surface gravity $\kappa \sim 0.6-1\  TeV^{-1}$
(this means $M\sim 5-10$ TeV when a=0), we obtain that, at
$\tilde{w}= \kappa/2\pi = T_H$, around the $20\%$ of the spectrum
comes from distances shorter than the new Planck length $l_{TeV}\sim
10^{-17} cm$, for $n=2-6$ and for spin zero particles. Moreover, at
frequencies $\tilde{w} \approx 3T_H$ the contribution of ultra-short
distances is of order of the total spectrum itself. For massless
$s=1/2$-particles the results can be obtained from the formulaes of
\cite{agullo-navarro-salas-olmo-parker} . In this case the
contribution from ultrashort distances is smaller than for spin zero
and it is around the $0.2\%$ of the spectrum at $\tilde{w}=
\kappa/2\pi = T_H$. In addition we find that, for $\kappa=0.9-1$,
and $n=6$ we need to go to frequencies $\tilde{w}\approx 5.5T_H$ and
$\tilde{w}\approx 5.6T_H$, respectively, to find short-distance
contributions of order of the fermionic thermal spectrum $(e^{2\pi
\tilde{w}\kappa^{-1}}+1)^{-1}$. For $\kappa=0.6-0.8$, and $n=2$ we
obtain $\tilde{w}\approx 6.2T_H$ and $\tilde{w}\approx 6.9T_H$,
respectively. Therefore, in TeV gravity scenarios the spectrum of
Hawking quanta is sensitive to transplanckian physics and
significant deviations from the thermal spectrum can emerge in the
``semiclassical'' phase of the evaporation.\\ {\bf
Acknowdelgements.} We thank L. Parker for collaboration on the topic
of this work. J. N-S also thanks M. Casals for  interesting
discussions on rotating black holes and TeV gravity. I. A. also
thanks R.M. Wald for useful discussions.

\end{document}